\documentclass{IEEEtran}

\usepackage{cite}
\usepackage{amsmath,amssymb,amsfonts}
\usepackage{algorithmic}
\usepackage{graphicx}
\usepackage{textcomp}
\usepackage{import}
\usepackage{amsfonts}
\usepackage{subeqnarray}
\usepackage{cases}
\usepackage{epstopdf}
\usepackage{xifthen}
\usepackage{pdfpages}
\usepackage{transparent}
\usepackage{wrapfig}
\usepackage{manfnt}
\usepackage{ragged2e}
\usepackage{mathtools}
\usepackage{xcolor}

\makeatother

\newtheorem{proposition}{Proposition}[section]

\newtheorem{remark}{Remark}[section]
\newtheorem{definition}{Definition}[section]

\newtheorem{assumption}{Assumption}[section]

\newenvironment{Proof}[1][Proof]{\textbf{#1.} }{\ \rule{0.5em}{0.5em}}

\newcommand{\xh}{\hat{x}}
\newcommand{\xt}{\tilde{x}}

\newcommand{\ut}{\tilde{u}}
\newcommand{\yt}{\tilde{y}}
 
\newcommand{\uh}{\hat{u}}
\newcommand{\yh}{\hat{y}}
\newcommand{\fhb}{\hat{f}_\partial}
\newcommand{\ehb}{\hat{e}_\partial}
\newcommand{\ftb}{\tilde{f}_\partial}
\newcommand{\etb}{\tilde{e}_\partial}
\newcommand{\Ht}{\tilde{H}}

\newcommand{\Ho}{\mathcal{H}}

\newcommand{\Bo}{\mathcal{B}}
\newcommand{\Co}{\mathcal{C}}

\newcommand{\fb}{f_\partial}
\newcommand{\eb}{e_\partial}
\newcommand{\zt}{(\zeta,t)}

\definecolor{blueMatlab}{rgb}{0, 0.4470, 0.7410}

\definecolor{orangeMatlab}{rgb}{0,0,0}
\definecolor{hector}{rgb}{0,0,0}

\begin{document}
\title{Infinite-dimensional observers for high order boundary-controlled port-Hamiltonian systems}
\author{Jesus-Pablo Toledo-Zucco, Yongxin Wu, Hector Ramirez, Yann Le Gorrec
\thanks{Manuscript received March 17, 2023; accepted May 3, 2023. The authors gratefully acknowledge the support of the EIPHI Graduate School (Contract: ANR-17-EURE-0002), ANR IMPACTS Project (Contract: ANR-21-CE48-0018) and ANID Basal Project FB0008 and FONDECYT 1231896.}
\thanks{J-P. Toledo-Zucco is with the Information Processing and Systems Department (DTIS), ONERA, the French Aerospace Lab, Toulouse 31000, France (email: jtoledoz@onera.fr).}
\thanks{H. Ramirez is with the Departamento Electrónica, Universidad Tecnica Federico Santa Maria, Valparaiso 2362735, Chile. (email: hector.ramireze@usm.cl)}
\thanks{Y. Wu and Y. Le Gorrec are with SUPMICROTECH, CNRS, FEMTO-ST, 25000 Besan\c{c}on, France (email: yongxin.wu@femto-st.fr; yann.le.gorrec@ens2m.fr).}
}

\maketitle

\begin{abstract}
This letter investigates the design of a class of infinite-dimensional observers for one dimensional (1D) boundary controlled port-Hamiltonian systems (BC-PHS)  defined by differential operators of order $N \geq 1$. The convergence of the proposed observer depends on the number and location of available boundary measurements. \textcolor{hector}{Asymptotic convergence is assured for $N\geq 1$, and provided that enough boundary measurements are available, exponential convergence can be assured for the cases $N=1$ and $N=2$.} Furthermore, in the case of partitioned BC-PHS with $N=2$, such as the Euler-Bernoulli beam, \textcolor{hector}{it is shown} that exponential convergence can be assured considering less available measurements. The Euler-Bernoulli beam model is used to illustrate the design of the proposed observers and to perform numerical simulations.  
\end{abstract}

%===============================================================================
\section{INTRODUCTION}\label{sec:Intro} 

Port Hamiltonian system (PHS) formulations \cite{Maschke1992ConferencePort} are widely used for the modeling and control design of complex multi-physical systems because their underlying structure arise from the intrinsic energy exchange between the sub-components of the physical system. This formalism has been used for the modeling of distributed parameter systems \cite{vanderSchaft2002JournalHamiltonian,Duindam2009BookModeling}, numerical spatial discretization \cite{Trenchant2018JournalFinite,Cardoso2020JournalAPartitioned} and from the definition of boundary controlled PHS (BC-PHS) to well-posedness and stability analysis \cite{LeGorrec2005JournalDirac,Jacob2012BookLinear}, as well as for control design \cite{Augner2014JournalStability,Ramirez2014JournalExponential,Macchelli2017JournalOnTheSynthesis,Macchelli2020JournalExponential}. Keeping in mind that these infinite dimensional systems are instrumented using a finite set of actuators and sensors, observer design is of key importance for this class of systems. This is even more the case for control design using state feedback. In this case the knowledge of the state variables of the infinite dimensional PHS and their initial conditions are required, implying that observer design for BC-PHS becomes a relevant and necessary task for practical control implementation, especially in the cases in which sensors are located at the boundaries of the system. 

The observer design for infinite dimensional (distributed parameter) systems is largely investigated in the literature. A survey on the topic can be found in \cite{Hidayat2011ConferenceObservers}. Generally speaking, the observer design for infinite dimensional systems is often treated in a case by case. For instance, the observer design for the wave equation has been investigated in  \cite{Guo2007JournalTheStabilization,Krstic2008JournalOutput,Guo2009JournalTheStrong,Smyshlyaev2009JournalBoundary,Meurer2011JournalTracking,Feng2016JournalObserver} and for the diffusion-convection-reaction processes in \cite{Smyshlyaev2005JournalBackstepping,Meurer2009JournalTracking}. It is not easy to get a general procedure for the observer design when dealing with infinite dimensional systems. 
%of a large class of linear distributed parameter systems (wave equation, Timoshenko beam, Euler-Bernoulli beam, etc.). 

In this letter, we investigate how to take advantage of the particular structure of BC-PHS for the observer design. In \cite{Toledo2020ConferencePassive,Malzer2020JournalStability,Wu2020JournalReduced,Toledo2020JournalObserver,Toledo2022ConferenceObserver} the observer design for BC-PHS has been investigated, however, the class of systems are restricted to PHS defined by first-order spatial differential operators. In the current contribution, a class of observer for higher-order differential operators subject to different boundary measurements and internal dissipation is proposed. 

The main contribution of this letter is summarized as follows:  infinite-dimensional observers for BC-PHSs defined by differential operators of order $N\geq1$ and internal \textcolor{orangeMatlab}{linear} dissipation are proposed in such a way that the error between the BC-PHS and the infinite-dimensional observer remains a BC-PHS. This allows to use existing results from the literature, in particular from \cite{Augner2014JournalStability}, to show the type of convergence depending on the available sensors. The proposed observers can be used for a large class of physical systems such as the wave equation, the Timoshenko and the Euler-Bernoulli beams, but also more complex systems arising from the interconnection of simple flexible structures.

This work extends the results proposed in \cite{Toledo2022ConferenceObserver} in which no internal dissipation was considered and the differential operator was limited to be of order one. In what follows, some simple conditions on the observer gains are provided to prove the asymptotic convergence of the observer in a quite general setting ($N\geq 1$ with possible dissipation) and similar conditions are provided to show the exponential convergence of the observer in the case of  differential operators of order up to $N=2$. The type of convergence of the proposed infinite-dimensional observers depends on passivity relations between the energy of the measurements and the energy flowing in/out through the spatial boundaries of the systems. 

The paper is organized as follows. Section \ref{sec:Pre_pro} gives some preliminaries on BC-PHS and the infinite-dimensional observer is defined. In Section \ref{Sec:General}, the observer design is shown and the different types of convergence are characterized in terms of the available boundary measurements. Section \ref{Section_Example} presents the clamped-free Euler-Bernoulli beam as illustrative example and in Section \ref{Sec:Simulation} numerical simulations are given. Finally Section \ref{Sec:ConPerspec} gives the final conclusion and lines of future work.

\section{Preliminaries and problem statement}\label{sec:Pre_pro}
We are interested in the design of infinite-dimensional observers for the following class of PDE
\begin{equation}\label{Eq:PDE}
\partial_t x(\zeta,t) = \sum_{k = 0}^N P_k \partial_\zeta^k(\Ho(\zeta) x(\zeta,t)), \,\,\, x(\zeta,0) = x_0(\zeta),
\end{equation}
where $\zeta \in [a,b]$ is the spatial variable and $t \geq0$ is the time, $x(\zeta,t) \in \textcolor{orangeMatlab}{\mathbb{R}}^n$ is the state variable with initial condition $x_0(\zeta)$, matrices $P_0 \in \textcolor{orangeMatlab}{\mathbb{R}}^{n\times n}$ and $P_k \in \textcolor{orangeMatlab}{\mathbb{R}}^{n\times n}$ with $k = \lbrace 1, \cdots , N \rbrace$ are such that $P_0^\top + P_0 \leq 0 $, $P_k^\top = (-1)^{k-1}P_k$, and we assume that $P_N$ is a non-singular matrix. \textcolor{orangeMatlab}{The Hamiltonian density matrix $\Ho (\zeta) \in \mathbb{R}^{n\times n}$ is a bounded and continuously differentiable matrix-valued function satisfying $\Ho (\zeta)=\Ho ^\top (\zeta)$ and $mI < \Ho (\zeta)< MI$ with $0< m < M$ for all $\zeta \in [a,b]$}. 
\textcolor{orangeMatlab}{\begin{remark}
Notice that, for simplicity and clarity of presentation, we restrict the state variable and parameters to belong to real spaces. However, as shown in \cite{Augner2014JournalStability}, the results can be \textcolor{hector}{extended} to state variables and parameters that are complex. An application case with complex variables and parameters is the Schrödinger equation (See \cite[Example~2.18]{Augner2014JournalStability}). 
\end{remark}}
The Hamiltonian of \eqref{Eq:PDE} is
\begin{equation}\label{Eq:H}
H(t) = \frac{1}{2} \int_a^b x^\ast(\zeta,t) \Ho(\zeta) x(\zeta,t)d \zeta, 
\end{equation}
The \textit{boundary port variables} \cite{LeGorrec2005JournalDirac} are defined as
\begin{equation}\label{D_Eq:BPV}
\begin{pmatrix}
\fb (t) \\
\eb (t)
\end{pmatrix}=\frac{1}{\sqrt{2}}\begin{pmatrix}
Q & -Q \\
I & I
\end{pmatrix}\begin{pmatrix}
\phi(b,t) \\
\phi(a,t) \\
\end{pmatrix},
\end{equation}
with
\begin{equation}
Q_{ij} = \begin{cases}
(-1)^{j-1}P_{i+j-1}, \quad i + j \leq N+1, \\
0, \quad\quad\quad\quad\quad\quad else,
\end{cases}
\end{equation}
and
\begin{equation}\label{Eq:PhiB}
\phi(b,t) = 
\begin{pmatrix}
\Ho(b) x(b,t) \\
{\partial_\zeta}(\Ho(b) x(b,t)) \\
\vdots \\
{\partial_\zeta^{N-1}}(\Ho(b) x(b,t))
\end{pmatrix}
\end{equation}
and similarly for $\phi(a,t)$. The input $u(t)$ and output $y(t)$ are defined as a linear combination of the boundary port variables
\begin{align}
u(t) = W_\Bo \left(\begin{smallmatrix}
\fb(t) \\
\eb(t) %Matrix0131K
 \\
\end{smallmatrix}\right)  ,  \label{Eq:Input}\\ 
y(t) = W_\Co \left(\begin{smallmatrix}
\fb(t) \\
\eb(t) \\
\end{smallmatrix}\right), \label{Eq:Output}
\end{align}
where $W_\Bo$ and $W_\Co$ are full rank matrices of size $Nn \times 2Nn$ such that  the following relations are satisfied $W_\Bo \Sigma W_\Bo^\top = 0$, $W_\Co \Sigma W_\Co^\top = 0$ and $W_\Co \Sigma W_\Bo^\top = I$, \textcolor{orangeMatlab}{with $\Sigma = \left( \begin{smallmatrix} 0 & I \\ I & 0 \end{smallmatrix} \right) \in \mathbb{R}^{2n \times 2n}$}. The system \eqref{Eq:PDE}, \eqref{Eq:Input}, \eqref{Eq:Output} is then a BC-PHS and its energy balance is
\begin{equation}\label{Eq:Hdot}
\dot{H}(t) = \frac{1}{2}\int_a^b {e}\zt \left(P_0^\top + P_0\right){e} \zt d\zeta + u(t)^\top y(t),
\end{equation}
with ${e} \zt := \Ho(\zeta) x \zt$ the effort variable. We are interested in the design of infinite-dimensional observers for this system. We assume that the input $u(t)$ is measured and that the power conjugated output $y(t)$ is partially measurable. We define the measured output as
\begin{equation}\label{Eq:MeasuredOutput}
y_m(t) = C_m y(t),
\end{equation}
with $C_m \in \mathbb{R}^{q \times n}$ and $q \leq n$.

In \cite{Toledo2022ConferenceObserver}, we proposed an observer design method for the undamped case and differential operator of order $N=1$. In the present work, we propose infinite dimensional observer designs for $N \geq 1$ and potential dissipation i.e. $P_0^\top + P_0 \leq 0$.

The considered class of observers is given in the following definition.  
\begin{definition}\label{Def:Observer}
The system 
\begin{equation}\label{Eq:Observer}
\begin{cases}
{\partial_t}\xh \zt =  \sum_{k = 0}^N P_k {\partial_\zeta^k} (\Ho \xh\zt) ,\, \xh (\zeta,0) = \xh_0 (\zeta),\\
 \uh(t) = W_\Bo \bigl(\begin{smallmatrix}
\fhb(t) \\
\ehb(t)
\end{smallmatrix}\bigr),     \\
 \yh(t)  = W_\Co \bigl(\begin{smallmatrix}
\fhb(t) \\
\ehb(t)
\end{smallmatrix}\bigr), \quad
\yh_m (t) = C_m \yh (t),
\end{cases}\hspace{-1cm}
\end{equation}
is a BC-PH observer for the system defined by \eqref{Eq:PDE}, \eqref{Eq:Input}, \eqref{Eq:Output}, \eqref{Eq:MeasuredOutput} if $\hat{x}(\zeta,t)$ converges to $x \zt$ for some initial condition $\xh_0(\zeta) \in L_2(a,b;\mathbb{R}^n) $ different from $x_0(\zeta)$. The boundary port variables $\left(\begin{smallmatrix}
\fhb(t) \\ \ehb(t)
\end{smallmatrix}\right)$ are defined as in \eqref{D_Eq:BPV} and \eqref{Eq:PhiB} with $\hat{x}\zt$ instead of $x\zt$.
\end{definition}

Since $u(t)$ and $y_m(t)$ are measured the observer input is designed as
\begin{equation}\label{Eq:InputObserver}
\uh(t) = u(t) + C_m^\top L (y_m(t) - \yh_m(t)), 
\end{equation}
with $L\in \mathbb{R}^{q \times q}$ such that $L + L^\top > 0$. The objective is then to characterize sufficient conditions on the available measurements in terms of $C_m$ and the observer gain $L$ such that the observer \eqref{Eq:Observer} with input \eqref{Eq:InputObserver} is an infinite-dimensional observer according to Definition \ref{Def:Observer}. \textcolor{hector}{Notice that, different from observers for linear ODEs in which the gain is generally a rectangular matrix, in this case the observer gain $L$ is a square matrix since it acts at the boundary and not on the domain of the PDE.} The error between the state of the plant and the observer is defined as $\xt\zt:= x\zt-\xh\zt$.
%\begin{equation}\label{Eq:Error}
%\xt\zt:= x\zt-\xh\zt.
%\end{equation}
The error system can then be written as the BC-PHS
\begin{equation}\label{Eq:ErrorSys}
\begin{cases}
\partial_t \xt\zt = \sum_{k = 0}^N P_k {\partial_\zeta^k} (\Ho \xt\zt),\, \xt(\zeta,0) = \xt_0(\zeta),\\
\ut(t) = W_\Bo \bigl(\begin{smallmatrix}
\ftb(t) \\
\etb(t)
\end{smallmatrix}\bigr), \\
\yt(t)  = W_\Co \bigl(\begin{smallmatrix}
\ftb(t) \\
\etb(t)
\end{smallmatrix}\bigr), \; \yt_m (t) = C_m \yt (t).
\end{cases}\hspace{-1cm}
\end{equation}
The Hamiltonian of the error system is defined in term of the state error as follows
\begin{equation}\label{Eq:HamiltonianError}
\Ht (t) :=  \dfrac{1}{2} \int_a^b {\xt^\ast \zt \Ho(\zeta) \xt \zt} d\zeta.
\end{equation}
and one can verify the following balance equation
\begin{equation}\label{Eq:ErrorBalance}
\dot{\Ht} (t) = \frac{1}{2}\int_a^b \tilde{e}\zt \left(P_0^\top + P_0\right)\tilde{e} \zt d\zeta + \ut (t)^\top \yt(t),
\end{equation}
where $\tilde{e}\zt : = \Ho (\zeta) \xt \zt$. \textcolor{hector}{Replace $\ut = u - \uh$ from \eqref{Eq:InputObserver} in \eqref{Eq:ErrorBalance} and \textcolor{hector}{since $P_0^\top + P_0 \leq 0$} it is obtained that
\begin{equation}\label{H_error_derive}
\dot{\Ht}\leq \ut^\top \yt = -\yt^\top C_m^\top L^\top C_m \yt = -\tfrac{1}{2}\yt_m^\top (L^\top + L) \yt_m,
\end{equation}
\textcolor{hector}{where we have used the properties of the quadratic vector product and that $L + L^\top > 0$. Since} $L^\top + L>0$ the error system \eqref{Eq:ErrorSys} converges to the origin and the observer system \eqref{Eq:Observer} qualifies as an infinite-dimensional observer according to Definition \ref{Def:Observer}. Furthermore, as observed in \eqref{H_error_derive} the rate of convergence explicitly depends on the observer gain $L$. In general the decay of \eqref{H_error_derive} is faster, and hence also the convergence of the observer, as $L$ grows bigger until certain value after which the system becomes over-damped \cite{Macchelli2017JournalOnTheSynthesis}.}

%%%%%%%%%%%%%%%%%%%%%%%OBSERVERS
\section{Observer design}\label{Sec:General}

In this section the different classes of observers and the type of convergence are presented according to the order of the differential operator of \eqref{Eq:PDE}. {In all cases the fundamental conditions for achieving convergence is the capability of the observer to bound the energy flowing through the boundary of the error system. As discussed in \cite{Ramirez2014JournalExponential} this is, roughly speaking, related to the passivity of  BC-PHS and to the definition of its inputs and outputs. The conditions that the observer gains have to satisfy to ensure the observer convergence are derived by applying the stability conditions presented in \cite{Augner2014JournalStability} to the error system.}

\subsection{{Asymptotic convergence: case $N > 1$}.}
In Proposition~\ref{Prop:Asymptotic} we propose simple conditions to check for the design of an observer with  asymptotic convergence to the plant system state.
\begin{proposition}\label{Prop:Asymptotic}
The system \eqref{Eq:Observer}-\eqref{Eq:InputObserver} is an observer according to Definition \ref{Def:Observer} with asymptotic convergence if there exists $\kappa>0$ such that
\begin{align*}
& \tfrac{1}{2} \yt_m^\top  \left(L^\top+ L \right) \yt_m  & \geq \kappa \sum_{k = 0}^{N-1} \Vert {\partial_\zeta^k}\left( \Ho \xt \right)(a) \Vert ^2, \\
& \qquad \mbox{or}  & \geq \kappa \sum_{k = 0}^{N-1} \Vert {\partial_\zeta^k}\left( \Ho \xt \right)(b) \Vert ^2 ,
\end{align*}
holds.
\end{proposition}
\begin{Proof}
\textcolor{hector}{Take the Hamiltonian error \eqref{Eq:HamiltonianError} as Lyapunov function and} since by \textcolor{orangeMatlab}{assumption} $-\tfrac{1}{2}\yt_m^\top (L^\top + L) \yt_m \leq -\kappa \sum_{k = 0}^{N-1} \Vert \partial_\zeta^k\left( \Ho \xt \right)(a) \Vert ^2 \; \mbox{(or $\zeta = b$),}
$
we have that
$
\dot{\Ht} \leq -\kappa \sum_{k = 0}^{N-1} \Vert {\partial_\zeta^k}\left( \Ho \xt \right)(a) \Vert ^2 \; \mbox{(or $\zeta = b$)}.
$
Using \cite[Proposition~2.11]{Augner2014JournalStability} we conclude that the error system converges to zero asymptotically.
\end{Proof}

\vspace{-0.4cm}
\subsection{Exponential convergence: case $N= 1$. }\label{Sec:ParticularN1}
\vspace{-0.1cm}
\begin{proposition}\label{Prop:N1Exponential}
The system \eqref{Eq:Observer}-\eqref{Eq:InputObserver} is an observer according to Definition \ref{Def:Observer} with exponential convergence if there exists $\kappa>0$ such that
\begin{align*}
\vspace{-0.6cm}
& \tfrac{1}{2} \yt_m^\top  \left(L^\top+ L \right) \yt_m  & \geq \kappa \Vert\Ho \xt (a) \Vert ^2, \\
& \qquad \mbox{or}  & \geq \kappa \Vert\Ho \xt (b) \Vert ^2, 
\end{align*} 
%\[
%\dfrac{1}{2} \yt_m^\top  \left(L^\top+ L \right) \yt_m \geq \kappa \Vert\Ho \xt \big|_{\zeta=a} \Vert ^2,
%\]
%or 
%\[
%\dfrac{1}{2} \yt_m^\top  \left(L^\top+ L \right) \yt_m \geq \kappa \Vert \Ho \xt \big|_{\zeta=b} \Vert ^2,
%\]
holds.
\end{proposition}
\begin{Proof}
Similarly to the proof of Proposition~\ref{Prop:Asymptotic} we use the Hamiltonian error functional as Lyapunov function.  The proof follows directly from \cite[Theorem~III.2]{Villegas2009JournalExponential} or \cite[Proposition~2.12]{Augner2014JournalStability}.
\end{Proof}
%\begin{remark}
%Proposition~\ref{Prop:N1Exponential} has been partially reported in \cite{Toledo2022ConferenceObserver} without considering internal dissipation, i.e., $P_0^\top + P_0 = 0$, and applied to a one-dimensional wave equation without dissipation. 
%\end{remark}

\subsection{Exponential convergence: case $N = 2$}\label{Sec:ParticularN2}

\begin{proposition}\label{Prop:N2Exponential}
The system \eqref{Eq:Observer}-\eqref{Eq:InputObserver} is an observer according to Definition \ref{Def:Observer} with exponential convergence if there exists $\kappa>0$ such that 
%$\tfrac{1}{2} \yt_m^\top  \left(L^\top+ L \right) \yt_m$ is greater or equal than one of the following expressions:
\begin{align*}
&\dfrac{1}{2} \yt_m^\top  \left(L^\top+ L \right) \yt_m \geq  \\
&\qquad \kappa \left( \Vert (\Ho \xt)(a) \Vert ^2  +  \Vert \partial_\zeta ( \Ho \xt )(a)  \Vert ^2  +  \Vert  (\Ho \xt)(b) \Vert ^2 \right),\\
&\mbox{or} \quad \kappa \left( \Vert (\Ho \xt)(a)  \Vert ^2 + \Vert \partial_\zeta (\Ho \xt)(a)  \Vert ^2  +  \Vert  \partial_\zeta (\Ho \xt)(b)\Vert ^2 \right), \\
&\mbox{or} \quad  \kappa \left( \Vert (\Ho \xt)(b)  \Vert ^2  +  \Vert \partial_\zeta ( \Ho \xt ) (b)  \Vert ^2  +  \Vert  (\Ho \xt)(a) \Vert ^2 \right), \\
&\mbox{or} \quad  \kappa \left( \Vert (\Ho \xt)(b)  \Vert ^2  +  \Vert \partial_\zeta ( \Ho \xt ) (b)  \Vert ^2  +  \Vert  \partial_\zeta (\Ho \xt) (a) \Vert ^2 \right),
\end{align*}
holds.
\end{proposition}
\begin{Proof}
Similarly to the proof of Proposition~\ref{Prop:Asymptotic} and taking the Hamiltonian error functional as Lyapunov function it is obtained that $\dot{\Ht}\leq  -\tfrac{1}{2}\yt_m^\top (L^\top + L) \yt_m$. Then if any of the conditions of Proposition~\ref{Prop:N2Exponential} holds and by direct application of \cite[Proposition~2.14]{Augner2014JournalStability}, the error system converges to zero exponentially. 
\end{Proof}

It is possible to relax the assumptions on the boundary dissipation of the system if the structure of the BC-PHS satisfies the following assumption.
%we present a relaxation of the previous proposition. In this case, by adding some constraint on the structure of the infinite-dimensional port-Hamiltonian system, the inequality conditions from Proposition~\ref{Prop:N2Exponential} can be relaxed.
\begin{assumption}\label{Assumption_structure}
We consider the system \eqref{Eq:PDE}, \eqref{Eq:Input}, \eqref{Eq:Output}, \eqref{Eq:MeasuredOutput} with $N = 2$. Assume $n$ an even number and that the state vector is split as $x\zt = (x_1\zt ,x_2 \zt) $ and the matrices $P_1$, $P_2$ and $\Ho(\zeta)$ such that
\[
P_1 = 
\begin{pmatrix} 
0 & Q_1 \\ 
Q_1 & 0 
\end{pmatrix}, \qquad 
P_2 = 
\begin{pmatrix} 
0 & - Q_2 
\\ 
Q_2 & 0 
\end{pmatrix}, 
\]
\[ 
\Ho(\zeta) = 
\begin{pmatrix} 
\Ho_1(\zeta) &  0 \\  
0& \Ho_2(\zeta) \end{pmatrix},
\] 
with $Q_1 \in \mathbb{R}^{n/2 \times n/2 }$ and $Q_2 \in \mathbb{R}^{n/2 \times n/2 }$ both self-adjoint matrices, and $Q_2$ invertible. $\Ho_1(\zeta)$ and $\Ho_2(\zeta)$ are uniformly positive matrices for all $\zeta$.
\end{assumption}
\begin{proposition}\label{Prop:N2ExponentialRelaxed}
Under Assumption~\ref{Assumption_structure}, the system \eqref{Eq:Observer}-\eqref{Eq:InputObserver} is an observer according to Definition \ref{Def:Observer} with exponential convergence if there exists $\kappa>0$ such that 
\begin{align*}
&\dfrac{1}{2} \yt_m^\top  \left(L^\top + L \right) \yt_m \geq  \\
&\qquad \kappa \left( \Vert (\Ho \xt)(a) \Vert ^2  +  \Vert \partial_\zeta ( \Ho_1 \xt_1 )(a)  \Vert ^2  +  \Vert  (\Ho_1 \xt_1)(b) \Vert ^2 \right),\\
&\mbox{or} \quad \kappa \left( \Vert (\Ho \xt)(a)  \Vert ^2 + \Vert \partial_\zeta (\Ho_2 \xt_2)(a)  \Vert ^2  +  \Vert  \partial_\zeta (\Ho_1 \xt_1)(b)\Vert ^2 \right), \\
&\mbox{or} \quad  \kappa \left( \Vert (\Ho \xt)(b)  \Vert ^2  +  \Vert \partial_\zeta ( \Ho_1 \xt_1 ) (b)  \Vert ^2  +  \Vert  (\Ho_1 \xt_1)(a) \Vert ^2 \right), \\
&\mbox{or} \quad  \kappa \left( \Vert (\Ho \xt)(b)  \Vert ^2  +  \Vert \partial_\zeta ( \Ho_2 \xt_2 ) (b)  \Vert ^2  +  \Vert  \partial_\zeta (\Ho_1 \xt_1) (a) \Vert ^2 \right),
\end{align*}
holds.
\end{proposition}
\begin{Proof}
The exponential convergence of the error system follows from the application of  \cite[Proposition~2.19]{Augner2014JournalStability}) considering that one of the conditions of Proposition~\ref{Prop:N2ExponentialRelaxed} hold.
\end{Proof}
\begin{remark}
Proposition~\ref{Prop:N2ExponentialRelaxed} is a special case of Proposition~\ref{Prop:N2Exponential} with some specific requirements on the matrices $P_1$ and $P_2$. The practical implication is that under these conditions some sensors can be removed and exponential convergence of the observer is still achieved. The Euler-Bernoulli beam fits perfectly into this specific structure. 
\end{remark}
%%%%%%%%%%%%%%%%%%%%%EXAMPLES

\section{Example:the Euler-Bernoulli beam}\label{Section_Example}

Consider the Euler-Bernoulli beam
\begin{equation}\label{Eq:EulerBernoulliPDE}
\rho(\zeta) \partial_t^2 w +d \partial_t w \zt= -\partial_\zeta^2 \left( EI(\zeta) {\partial_\zeta^2}w \zt \right),
\end{equation}
with $\zeta \in [0,1]$ and $t\geq 0$. $\rho(\zeta) > 0 $ is the mass density, $E>0$ is the elastic modulus, $I(\zeta)>0$ is the second moment of area of the cross section and $d>0$ the internal damping coefficient. $w \zt $ is the deflection of the beam. The initial conditions are defined as $w(\zeta,0) = w_0(\zeta)$ and $\tfrac{\partial w}{\partial t}(\zeta,0) = v_0(\zeta)$. Defining the state variables as
\begin{equation}
x_1 \zt = \rho (\zeta) {\partial_t} w \zt , \quad x_2 \zt = {\partial_\zeta^2} w \zt
\end{equation}
the PDE \eqref{Eq:EulerBernoulliPDE} can then be written as a BC-PHS with $N=2$ and
\begin{equation}\label{Eq:Matrices}
\begin{split}
P_0 &=\begin{pmatrix}
 -d & 0 \\ 0 & 0
\end{pmatrix}, \quad P_1 =\begin{pmatrix}
 0 & 0 \\ 0 & 0
\end{pmatrix},  P_2 =\begin{pmatrix}
 0 & -1 \\ 1 & 0
\end{pmatrix},\\
\Ho &=\begin{pmatrix}
 \tfrac{1}{\rho(\zeta)} & 0 \\ 0 & EI(\zeta)
\end{pmatrix} =: \begin{pmatrix}
\Ho_1(\zeta) & 0 \\ 0 & \Ho_2(\zeta)
\end{pmatrix},
\end{split}
\end{equation}
The inputs and outputs are defined by
\begin{equation} \label{Eq:InputBeam}
\begin{matrix}
u(t) = & y(t) = \\
\begin{pmatrix}
\Ho_1x_1(0,t) \\
{\partial_\zeta} \left(\Ho_1 x_1 \right)(0,t) \\
\Ho_2x_2(1,t) \\
{\partial_\zeta} \left(\Ho_2 x_2 \right)(1,t)
\end{pmatrix}, &
\begin{pmatrix}
{\partial_\zeta} \left( \Ho_2 x_2 \right)(0,t) \\
-\Ho_2 x_2(0,t) \\
{\partial_\zeta} \left( \Ho_1x_1 \right) (1,t)\\
-\Ho_1x_1(1,t)
\end{pmatrix}.
\end{matrix}
%\begin{equation} \label{Eq:InputBeam}
%\begin{matrix}
%u(t) = & y(t) = \\
%\begin{pmatrix}
%\frac{1}{\rho(0)}x_1(0,t) \\
%{\partial_\zeta} \left(\frac{1}{\rho(0)}x_1(0,t) \right) \\
%EI(1)x_2(1,t) \\
%{\partial_\zeta} \left( EI(1)x_2(1,t) \right)
%\end{pmatrix}, &
%\begin{pmatrix}
%{\partial_\zeta} \left( EI(0)x_2(0,t) \right) \\
%-EI(0)x_2(0,t) \\
%{\partial_\zeta} \left(\frac{1}{\rho(1)}x_1(1,t) \right) \\
%-\frac{1}{\rho(1)}x_1(1,t)
%\end{pmatrix}.
%\end{matrix}
%u(t) = 
%\begin{pmatrix}
%\frac{1}{\rho(0)}x_1(0,t) \\
%\dfrac{\partial}{\partial \zeta} \left(\frac{1}{\rho(0)}x_1(0,t) \right) \\
%EI(1)x_2(1,t) \\
%\dfrac{\partial}{\partial \zeta} \left( EI(1)x_2(1,t) \right)
%\end{pmatrix}, 
%\end{equation}
%\begin{equation}
%y(t) =\begin{pmatrix}
%\dfrac{\partial}{\partial \zeta} \left( EI(0)x_2(0,t) \right) \\
%-EI(0)x_2(0,t) \\
%\dfrac{\partial}{\partial \zeta} \left(\frac{1}{\rho(1)}x_1(1,t) \right) \\
%-\frac{1}{\rho(1)}x_1(1,t)
%\end{pmatrix}.
\end{equation}
The Hamiltonian 
\[ H(t) = \dfrac{1}{2} \int_a ^b \begin{pmatrix} x_1  \\ x_2  \end{pmatrix} ^\top \begin{pmatrix}
\Ho_1(\zeta) & 0 \\ 0 & \Ho_2(\zeta)
\end{pmatrix}\begin{pmatrix} x_1  \\ x_2 \end{pmatrix} d\zeta \]
then satisfies $ \dot{H}(t) = - \int_a^b d \left(\Ho_1(\zeta) x_1 \zt \right)^2d\zeta + u(t)^\top y(t).$

\vspace{-0.5cm}
\subsection{BC-PHS observer}\label{Subsec:Observer}
The infinite-dimensional observer \eqref{Eq:Observer} is given by
\begin{equation*}%\label{Eq:ObserverBeam}
%\begin{split}
{\partial_t}
\begin{pmatrix} 
\xh_1 \\ 
\xh_2
\end{pmatrix} = 
\begin{pmatrix} 
0 & -1 \\ 
1 & 0 
\end{pmatrix}
{\partial_\zeta^2}
\begin{pmatrix} 
\Ho_1 \xh_1 \\ 
\Ho_2 \xh_2 
\end{pmatrix} + 
\begin{pmatrix} 
-d & 0 \\ 
0 & 0
\end{pmatrix}
\begin{pmatrix} 
\Ho_1 \xh_1 \\ 
\Ho_2 \xh_2 
\end{pmatrix}, 
%\begin{array}{l} \xh_1 (\zeta,0) = \xh_{10} (\zeta)\\ \xh_2(\zeta,0)   = \xh_{20} (\zeta)\end{array} 
%\end{split}
\end{equation*}
with $\xh_1 (\zeta,0) = \xh_{10} (\zeta)$, $\xh_2(\zeta,0) = \xh_{20} (\zeta)$ and
\begin{equation}\label{Eq:InputBeamObserver}
\uh = 
\begin{pmatrix}
\Ho_1 \xh_1 (0) \\
{\partial_\zeta} \left(\Ho_1\xh_1 \right)(0) \\
\Ho_2 \xh_2(1) \\
{\partial_\zeta} \left( \Ho_2 \xh_2 \right) (1)
\end{pmatrix}, \,
 \yh = 
\begin{pmatrix}
{\partial_\zeta} \left( \Ho_2 \xh_2 \right) (0) \\
-\Ho_2 \xh_2 (0)\\
{\partial_\zeta} \left(\Ho_1 \xh_1\right) (1) \\
-\Ho_1 \xh_1 (1) 
\end{pmatrix}.
\end{equation}
%We shall now investigate the convergence properties of the proposed observer \eqref{Eq:Observer}-\eqref{Eq:InputObserver} regarding the available measurements. 
{Since the differential operator is of size $N=2$, Proposition~\ref{Prop:Asymptotic} and Proposition~\ref{Prop:N2Exponential} can be used to guarantee, respectively, asymptotic and exponential convergence depending on the available measurements. Moreover, in this example, by employing Proposition~\ref{Prop:N2ExponentialRelaxed}, one can take advantage of the PDE structure in such a way that exponential convergence can be guaranteed using less sensors.}
\subsection{Three boundary measurements}\label{Example:three_measuremens}
Consider that three boundary measurements are available
\[ 
C_m = 
\begin{pmatrix}
0 & 1 & 0 & 0 \\
0 & 0 & 1 & 0 \\
0 & 0 & 0 & 1
\end{pmatrix}, \quad
y_m = 
\begin{pmatrix}
-\Ho_2 x_2 (0)\\
{\partial_\zeta} \left(\Ho_1 x_1\right) (1) \\
-\Ho_1 x_1 (1) 
\end{pmatrix},
\]
For simplicity we use a diagonal observer gain $L = \mbox{diag}(l_1, l_2, l_3)$, with $l_1,l_2,l_3>0$. To verify Proposition~\ref{Prop:N2Exponential} we first compute the left hand side of the inequality 
\begin{multline*}
\tfrac{1}{2} \yt_m^\top (L^\top + L )\yt_m =\\
l_1 \left( (-\Ho_2  \xt_2)(0)\right)^2 +  l_2 \left( \partial_\zeta (\Ho_1  \xt_1)(1)\right)^2 + l_3 \left( (-\Ho_1  \xt_1)(1)\right)^2.
\end{multline*}
From the third relation of Proposition~\ref{Prop:N2Exponential} and using the boundary condition  $\ut =-C_m^\top L C_m \yt$, we obtain
%\begin{align*}
%\Vert (\Ho \xt) \big|_{\zeta = 1}  \Vert ^2  &+  \Vert \partial_\zeta ( \Ho \xt ) \big|_{\zeta = 1}  \Vert ^2  +  \Vert  (\Ho \xt) \big|_{\zeta = 0}  \Vert ^2 \\
%& = \left((\Ho_1 \xt_1) \big|_{\zeta = 1}\right)^2 +   \left((\Ho_2 \xt_2) \big|_{\zeta = 1}\right)^2 \\
%& \quad + \left(\partial_\zeta(\Ho_1 \xt_1) \big|_{\zeta = 1}\right)^2 +  \left(\partial_\zeta(\Ho_2 \xt_2) \big|_{\zeta = 1}\right)^2 \\
%& \quad \quad + \left((\Ho_1 \xt_1) \big|_{\zeta = 0}\right)^2 +   \left((\Ho_2 \xt_2) \big|_{\zeta = 0}\right)^2.
%\end{align*}
%and replace the known values from $\ut = u- \uh$ which is written as
%\begin{equation*}
%\ut = \left(\begin{array}{r}
%\Ho_1 \xt_1 \big|_{\zeta = 0} \\
%{\partial_\zeta} \left(\Ho_1\xt_1 \right) \big|_{\zeta = 0} \\
%\Ho_2 \xt_2\big|_{\zeta = 1} \\
%{\partial_\zeta} \left( \Ho_2 \xt_2 \right) \big|_{\zeta = 1}
%\end{array}\right) = 
%\left(\begin{array}{r}
%0 \\
%-l_1(-\Ho_2 \xt_2) \big|_{\zeta = 0}\\
%-l_2 {\partial_\zeta} \left(\Ho_1 \xt_1\right) \big|_{\zeta = 1} \\
%-l_3 (-\Ho_1 \xt_1) \big|_{\zeta = 1}
%\end{array}\right).
%\end{equation*}
%Then, we obtain the condition $(iii)$ as follows:
\begin{multline*}
\Vert (\Ho \xt) (1)  \Vert ^2  +  \Vert \partial_\zeta ( \Ho \xt ) (1)  \Vert ^2  +  \Vert  (\Ho \xt) (1)  \Vert ^2 =  \\
(1+l_3^2)\left((\Ho_1 \xt_1) (1)\right)^2 + (1+l_2^2)\left(\partial_\zeta(\Ho_1 \xt_1) (1)\right)^2 \\ +   \left((\Ho_2 \xt_2) (0)\right)^2,
\end{multline*}
so it is always possible to find a $\kappa>0$ such that Proposition~\ref{Prop:N2Exponential} is satisfied.
\subsection{Two boundary measurements}\label{Example:two_meassurements}
Consider that only two measured outputs are available
\begin{equation}\label{two_meassurements}
C_m = 
\begin{pmatrix}
0 & 0 & 1 & 0 \\
0 & 0 & 0 & 1
\end{pmatrix}, \quad
y_m = 
\begin{pmatrix}
{\partial_\zeta} \left(\Ho_1 x_1\right)(1) \\
-\Ho_1 x_1 (1) 
\end{pmatrix}
\end{equation}
hence $q = 2$. We shall first investigate if the observer guarantees asymptotic stability using Proposition~\ref{Prop:Asymptotic}. Define $L = \mbox{diag}(l_1, l_2)$, with $l_1,l_2>0$, and compute the left hand side of the inequality 
\begin{multline}\label{left_hand_side}
\tfrac{1}{2} \yt_m^\top (L^\top + L )\yt_m \\
= l_1 \left( \partial_\zeta (\Ho_1  \xt_1) (1)\right)^2 + l_2 \left(  (-\Ho_1  \xt_1)(1)\right)^2.
\end{multline}
From the second condition of Proposition~\ref{Prop:Asymptotic} and using the boundary conditions $\ut = -C_m^\top L C_m \yt$ we obtain
% \begin{align*}
% \Ho_2 \xt_2 \big|_{\zeta=b} &= -l_1 \partial_\zeta (\Ho_1 \xt_1 \big|_{\zeta = b}), \quad and \\
%  \partial_\zeta ( \Ho_2 \xt_2 ) \big|_{\zeta=b}  &= -l_2. ( -\Ho_1 \xt_1 \big|_{\zeta = b} ) 
%  \end{align*} 
%   from the input $\ut = u - \uh$. Then, 
\begin{multline*}
\sum_{k = 0}^{1} \Vert \tfrac{\partial ^k}{\partial \zeta ^k}\left( \Ho \xt \right)(1) \Vert ^2 \\
= (1+l_1^2)\partial_\zeta (\Ho_1 \xt_1 (1))^2
  + (1+l_2^2)(\Ho_1 \xt_1 (1))^2
\end{multline*}
so it is always possible to find a $\kappa>0$ such that Proposition~\ref{Prop:Asymptotic} is satisfied. Furthermore, since the BC-PHS formulation of the Euler-Bernoulli beam satisfies Assumption~\ref{Assumption_structure}, $n=2$, $Q_1 = 0$ (scalar self-adjoint) and $Q_2 = 1$ (scalar self-adjoint and invertible), it is possible to use Proposition~\ref{Prop:N2ExponentialRelaxed} to investigate exponential convergence even if only two measurements are available. Consider again the measurements \eqref{two_meassurements} and the same observer matrix $L = \mbox{diag}({l_1}, {l_2})$, with $l_1,l_2>0$. 
%\[ 
%C_m = 
%\begin{pmatrix}
%0 & 0 & 1 & 0 \\
%0 & 0 & 0 & 1
%\end{pmatrix}, \quad
%y_m = 
%\begin{pmatrix}
%{\partial_\zeta} \left(\Ho_1 x_1\right) (1) \\
%-\Ho_1 x_1 (1) 
%\end{pmatrix}.
%\]
%and define $L = \mbox{diag}(\sqrt{l_1}, \sqrt{l_2})$, with $l_1,l_2>0$. Computing 
%\[
%\tfrac{1}{2} \yt_m^\top (L^\top + L )\yt_m = l_1 \left( \partial_\zeta (\Ho_1  \xt_1)(1)\right)^2 + l_2 \left(  (-\Ho_1  \xt_1)(1)\right)^2 
%\]
From the third relation of Proposition~\ref{Prop:N2ExponentialRelaxed} and using that $\ut =-C_m^\top L C_m \yt$ we have that 
%\begin{align*}
%\Vert (\Ho \xt) \big|_{\zeta = 1}  \Vert ^2  &+  \Vert \partial_\zeta ( \Ho_1 \xt_1 ) \big|_{\zeta = 1}  \Vert ^2  +  \Vert  (\Ho_1 \xt_1) \big|_{\zeta = 0}  \Vert ^2 \\
%& = \left((\Ho_1 \xt_1) \big|_{\zeta = 1}\right)^2 +   \left((\Ho_2 \xt_2) \big|_{\zeta = 1}\right)^2 \\
%& \quad + \left(\partial_\zeta(\Ho_1 \xt_1) \big|_{\zeta = 1}\right)^2  \\
%& \quad \quad + \left((\Ho_1 \xt_1) \big|_{\zeta = 0}\right)^2 .
%\end{align*}
%and replace the known values from $\ut = u- \uh$ which is written as
%\begin{equation*}
%\ut = \left(\begin{array}{r}
%\Ho_1 \xt_1 \big|_{\zeta = 0} \\
%{\partial_\zeta} \left(\Ho_1\xt_1 \right) \big|_{\zeta = 0} \\
%\Ho_2 \xt_2\big|_{\zeta = 1} \\
%{\partial_\zeta} \left( \Ho_2 \xt_2 \right) \big|_{\zeta = 1}
%\end{array}\right) = 
%\left(\begin{array}{r}
%0 \\
%0\\
%-l_2 {\partial_\zeta} \left(\Ho_1 \xt_1\right) \big|_{\zeta = 1} \\
%-l_3 (-\Ho_1 \xt_1) \big|_{\zeta = 1}
%\end{array}\right).
%\end{equation*}
%Then, we obtain the condition $(iii)$ as follows:
\begin{multline*}
\Vert (\Ho \xt) (1)  \Vert ^2  +  \Vert \partial_\zeta ( \Ho_1 \xt_1 ) (1) \Vert ^2  +  \Vert  (\Ho_1 \xt_1) (0)  \Vert ^2 = \\
\left((\Ho_1 \xt_1) (1) \right)^2 + (1+l_2^2)\left(\partial_\zeta(\Ho_1 \xt_1) (1) \right)^2, 
\end{multline*}
and comparing with \eqref{left_hand_side} we conclude that there is always a $\kappa >0$ such that Proposition~\ref{Prop:N2ExponentialRelaxed} is satisfied. Hence, because the structure of the Euler-Bernoulli satisfies the conditions of Assumption~\ref{Assumption_structure}, it is possible to assure exponential convergence of the observer with only two boundary measurements.

\section{Numerical simulations}
\label{Sec:Simulation}

In this section the performances of the observer proposed in Section \ref{Example:two_meassurements} are illustrated using numerical simulations considering, for simplicity, unitary parameters for the beam, i.e., $\zeta\in [0,1]$ and constant and unitary parameters associated to the energy storing elements, i.e., $\rho(\zeta) = 1$ and $EI(\zeta) = 1$, and damping term $d = 0.2$. For the spatial discretization structure preserving finite-differences on staggered grids \cite{Trenchant2018JournalFinite} is used with $n_d=140$ state variables. \textcolor{orangeMatlab}{Due to the stiffness of the PDEs, we use the $ode15$ environment from Matlab for the time discretization, obtaining less expensive and faster numerical simulation compared to, for instance, \textcolor{hector}{$ode45$}}. The beam model is simulated using as input $u(t) = 0$ and initial condition $x_0(\zeta) = \left[\begin{smallmatrix}0 \\ \zeta -0.9 \end{smallmatrix}\right]$, which represents the beam in equilibrium with a bending moment $EI(1)x_2(1,0) = 0.1 \, Nm$ and a shear force $\partial_\zeta(EI(1)x_2(1,0))= 1 \,N$. %\textcolor{red}{(Which one is force and which one is torque ?)}. 
The observer is simulated with the input \eqref{Eq:InputBeamObserver} using \eqref{Eq:InputObserver} with $L = \mbox{diag}(l_1,l_2)$ with design parameters $l_1 = 0.1$ and $l_2 = 1$. The initial conditions of the observer are $\hat{x}_{10}(\zeta) =\hat{x}_{20}(\zeta) = 0$.

\begin{figure}%[!h]
\begin{center}
\includegraphics[width=0.46\textwidth]{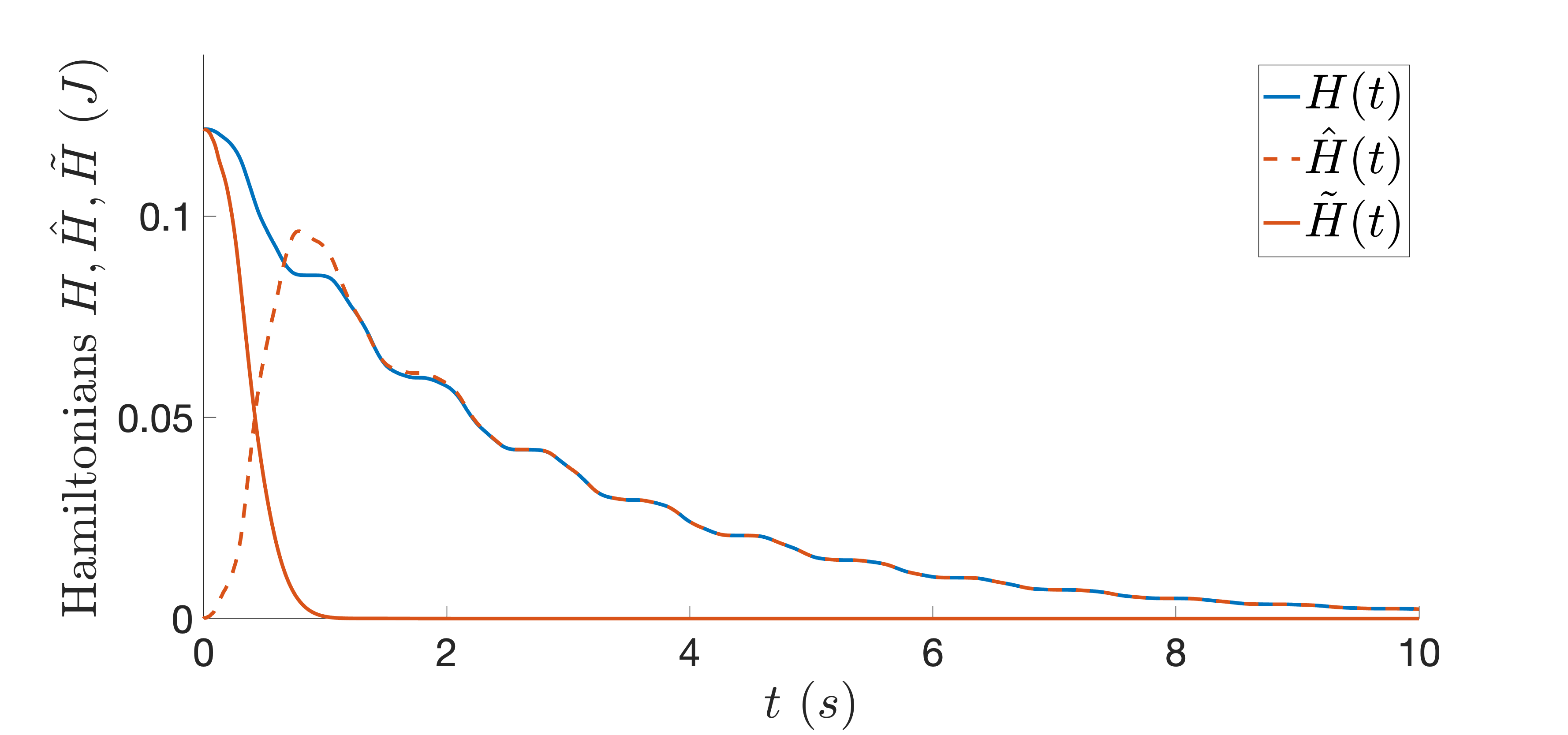}  
\caption{Hamiltonian of the system (solid blue), estimated one (dashed orange) and the Hamiltonian error (solid orange).} 
\label{Fig:Hamiltonian}                           
\end{center}
\end{figure}

\begin{figure}%[!h]
\begin{center}
\includegraphics[width=0.42\textwidth]{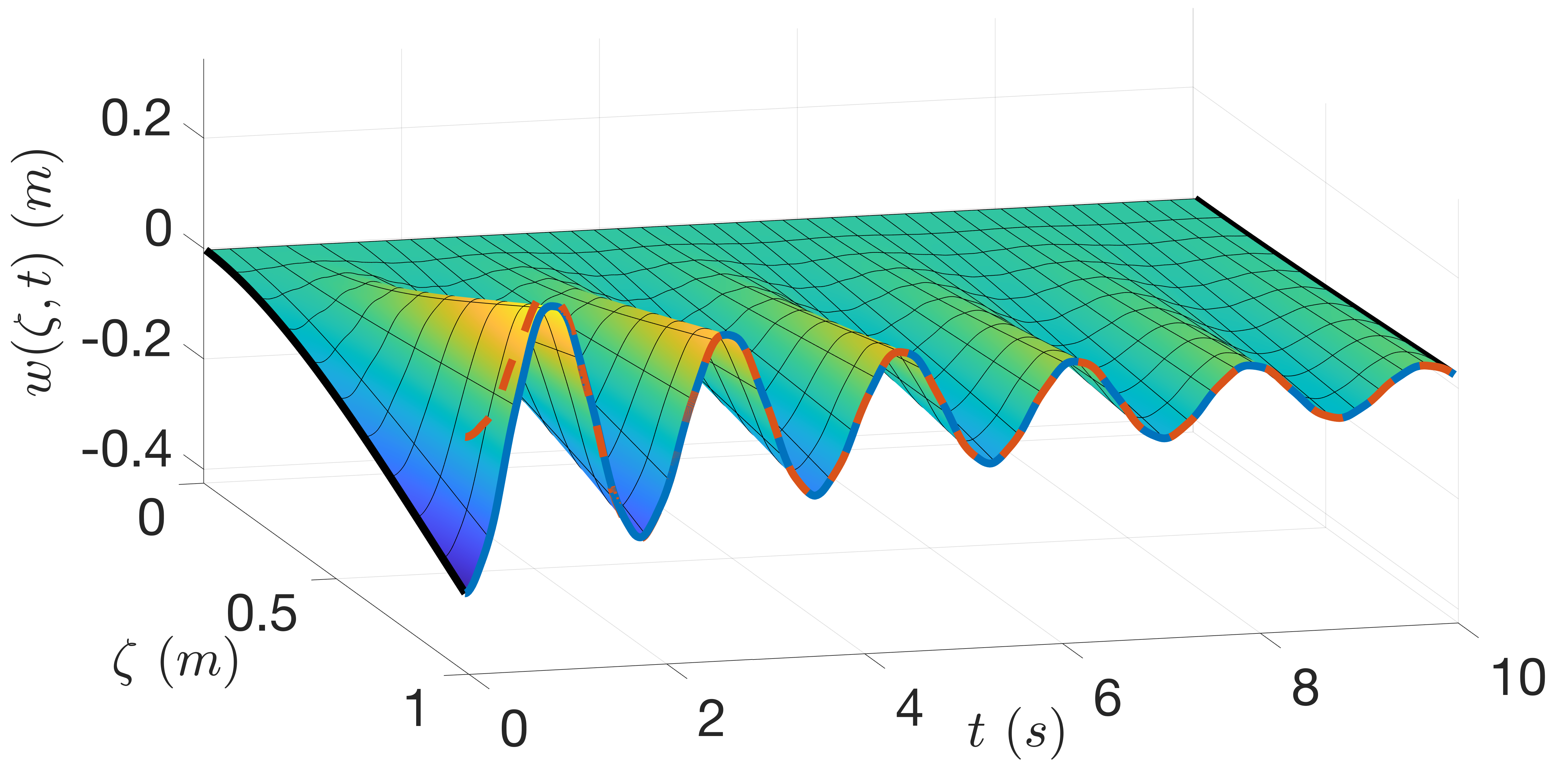}  
\caption{Beam deformation along time and space. The solid blue line is the end-tip position of the beam whereas the dashed orange line is the estimated one.} 
\label{Fig:Def}                           
\end{center}
\end{figure}

\begin{figure}%[!h]
\begin{center}
\includegraphics[width=0.42\textwidth]{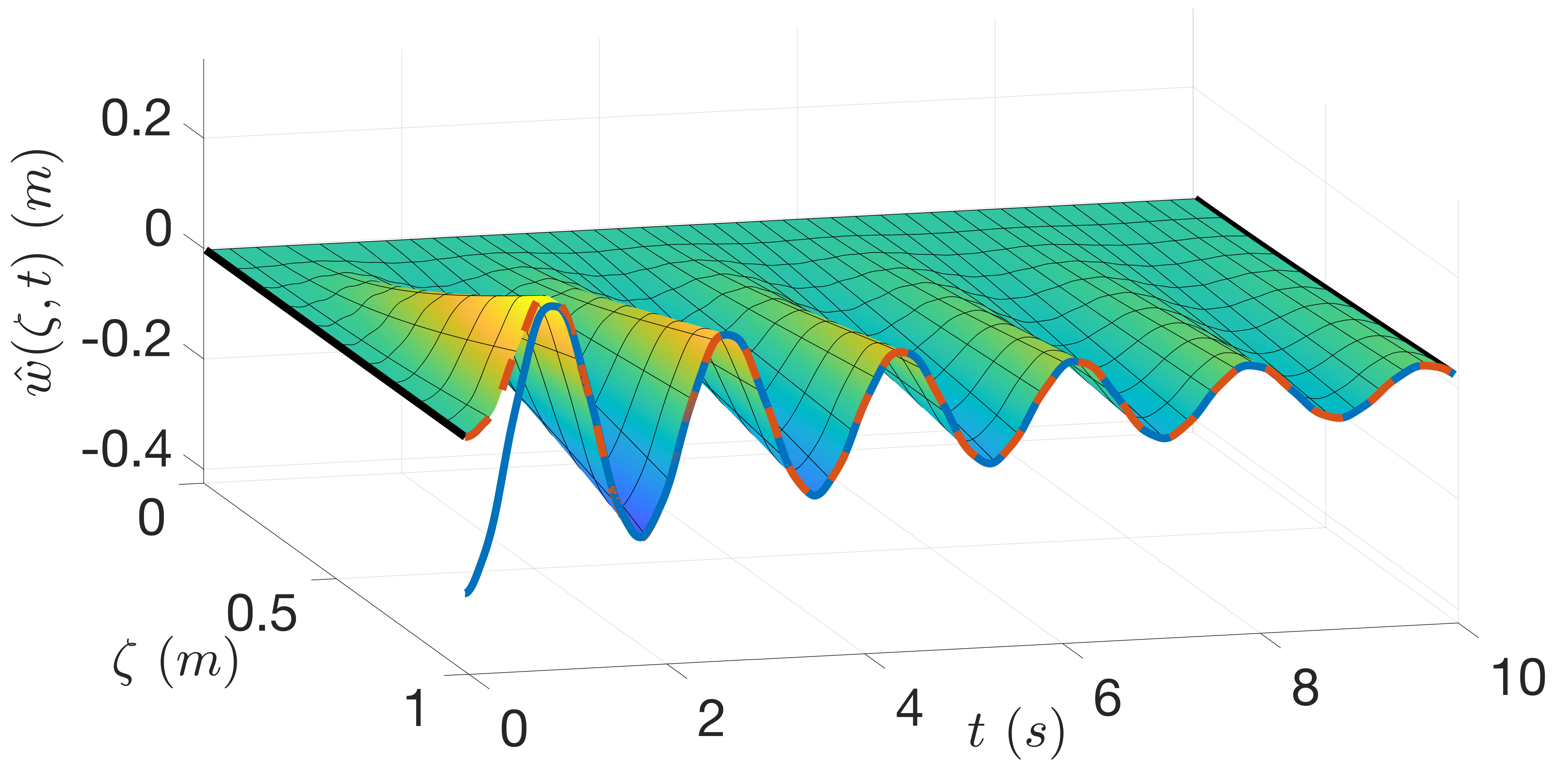}  
\caption{Estimation of the deformation of the beam.} 
\label{Fig:Def_h}                           
\end{center}
\vspace{-0.5cm}
\end{figure}

\begin{figure}%[!h]
\vspace{-0.5cm}
\begin{center}
\includegraphics[width=0.42\textwidth]{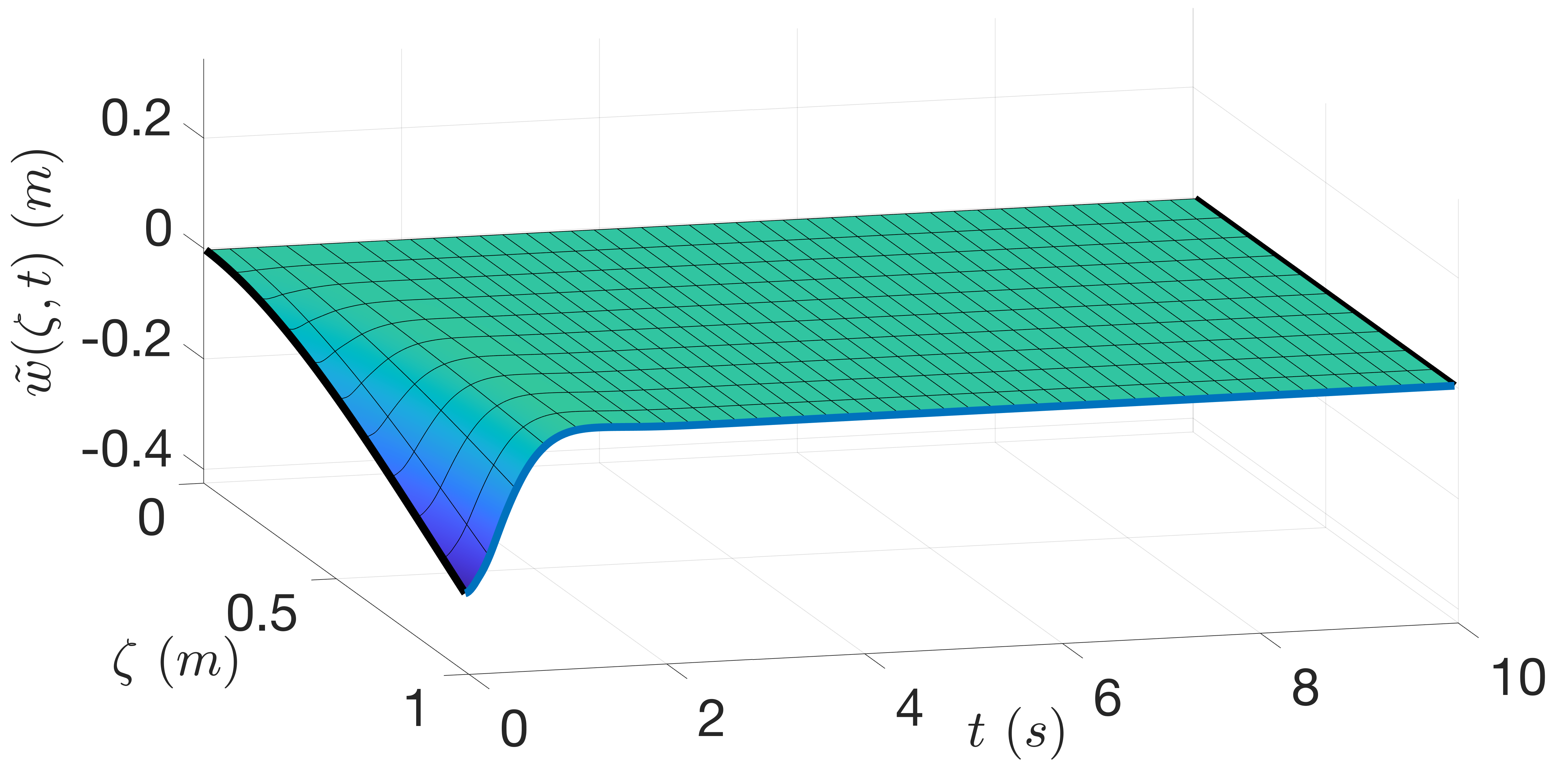}  
\caption{Error between the beam deformation and the estimated one. } 
\label{Fig:Def_t}                           
\end{center}
%\vspace{-0.5cm}
\end{figure}

 Fig. \ref{Fig:Hamiltonian} shows the system, observer and error system Hamiltonian, respectively, $H(t)$, $\hat{H}(t)$ and $\tilde{H}(t)$.
%These values are quadratic with respect to the discretized state, {\it i.e.,} $H(t) = \tfrac{h}{2}x_d(t)^\top Q_d x_d(t)$, $\hat{H}(t) = \tfrac{h}{2}\hat{x}_d(t)^\top Q_d \hat{x}_d(t)$, and $\tilde{H}(t) = \tfrac{h}{2}(x_d(t)-\hat{x}_d(t))^\top Q_d(x_d(t)-\hat{x}_d(t))$, where $h$ is the spatial discretization step and $Q_d$ is a diagonal matrix containing the values of $\rho(\zeta)$ and $EI(\zeta)$ scaled with respect to $h$. 
Since there is internal dissipation the Hamiltonian goes to zero for all the systems. The exponential convergence of the observer is appreciated in the response of $\tilde{H}(t)$. Fig. \ref{Fig:Def} shows the deformation of the beam along time and space. One can see that due to the internal dissipation, the beam deformation decreases as time increases. The solid blue line shows the end-tip position of the beam whereas the dashed orange line shows the estimated end-tip position. It is observed that around $t = 1 \, s$, the dashed orange line superposes the solid blue. Fig. \ref{Fig:Def_h} shows the estimation of the beam deformation. Starting from a zero initial condition the observer is able to accurately described the beam deformation around second $t = 1 \, s$. The error between the beam deformation and the estimated one is shown in Fig.~\ref{Fig:Def_t}.

%\newpage
\vspace{-0.4cm}
\subsection{Performance of the observer}\label{Sec:Perform}
\begin{figure}[!h]
\begin{center}
\includegraphics[width=0.4\textwidth]{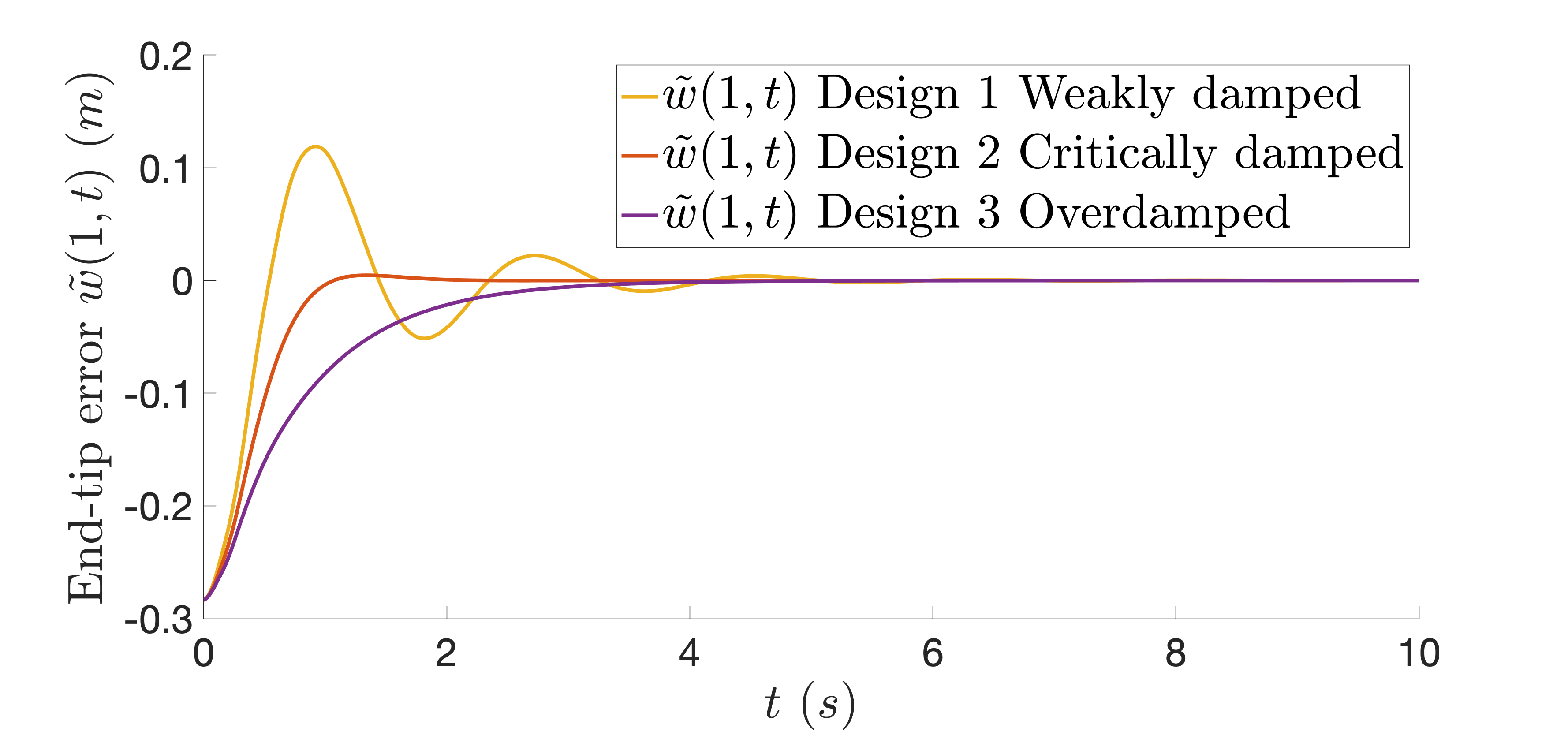}  
\caption{Error of the end-tip position of the beam for design 1 (yellow), design 2 (orange) and design 3 (violet).} 
\label{Fig:End_Tip_Error_Designs}                           
\end{center}
\vspace{-0.5cm}
\end{figure}

\begin{figure}[!h]
\begin{center}
\includegraphics[width=0.4\textwidth]{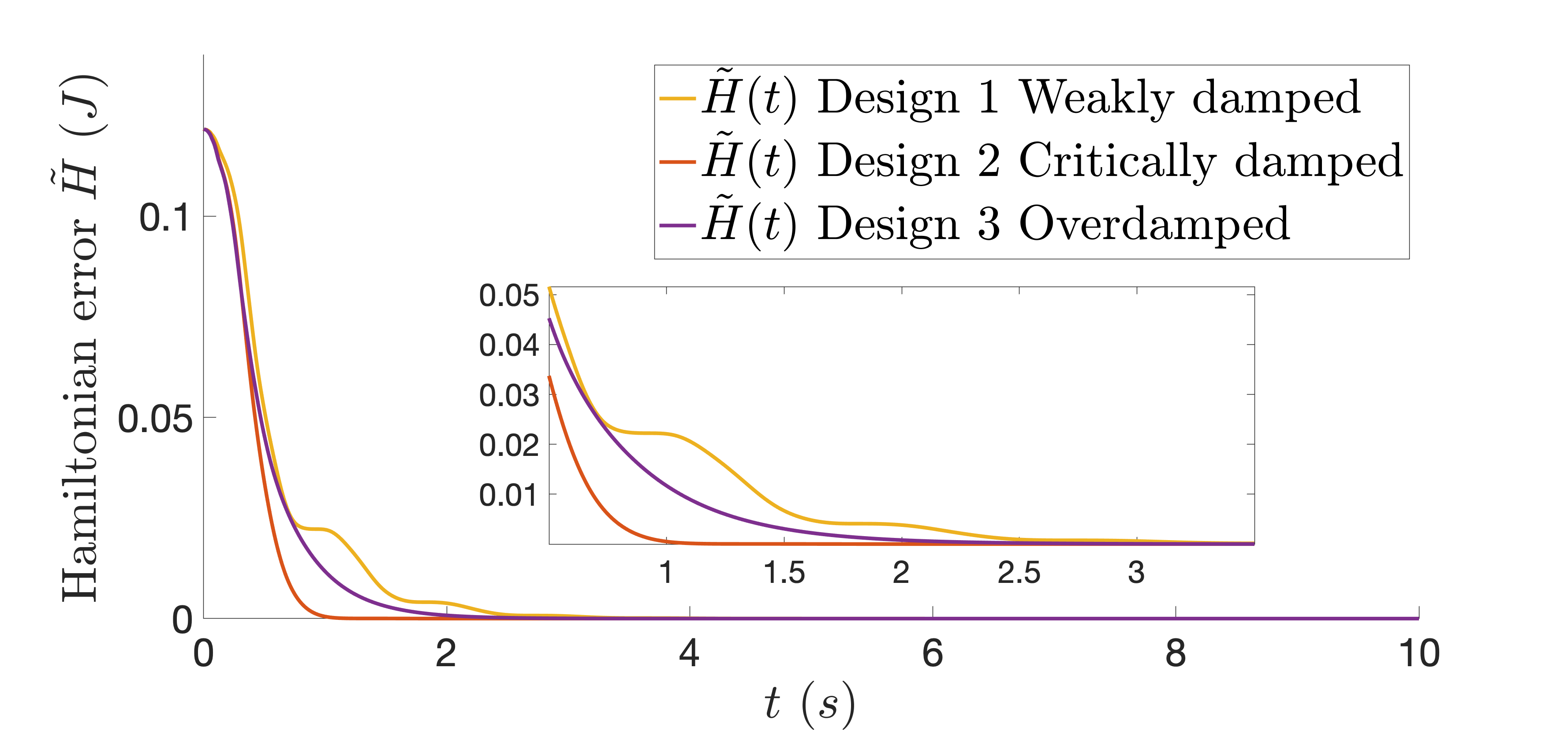}  
\caption{Hamiltonian error $\tilde{H}(t)$ for the weakly, critically and over damped cases.} 
\label{Fig:Energy_Designs}                           
\end{center}
\vspace{-0.5cm}
\end{figure}

The performance of the infinite-dimensional observer for different design parameters is commented. The dynamic of the error system behaves as a BC-PHS with a boundary damper. The damper term is proportional to the observer gain matrix $L$. The behavior of the error can hence be classified in three zones: $(i)$ {\it weakly damped}, $(ii)$ {\it critically damped} and $(iii)$ {\it overdamped}. Table \ref{Table:Design} gives different values of $L$ corresponding to each of the aforementioned cases.
\begin{table}[!h]
\vspace{-0.2cm}
\centering
\caption{Design parameteres.}
\label{Table:Design}
\begin{tabular}{||c c c l||} 
 \hline
Design & $l_1$ & $l_2$  & Performance\\ [0.5ex] 
 \hline\hline
 1 & 0.03 & 0.30 & Weakly damped  \\ 
 \hline
 2 & 0.10 & 1.00  & Critically damped\\
 \hline
 3 & 0.20 & 2.00  & Overdamped\\
 \hline
\end{tabular}
\end{table}
Fig. \ref{Fig:End_Tip_Error_Designs} shows the behavior of the error between the end-tip position of the beam and the estimated one for the three different designs. As for a second order system, it is appreciated that in the weakly damped case there is a big overshoot, in the critically damped case there is almost no overshoot and that in the overdamped case there are no oscillations and the time response is slower than the critically damped case.
Finally, Fig. \ref{Fig:Energy_Designs} shows the Hamiltonian error computed as in \eqref{Eq:HamiltonianError} for the three cases. We can see that the Hamiltonian error converges to zero faster for the critically damped. For this design the critically damped Hamiltonian error converges to zero in around $t = 1\, s$, whereas for the weakly damped case, the convergence is around $t = 3 \,s$ and for the overdamped case around $t = 2 \, s$.

\vspace{-0.5cm}
\section{Conclusion and future work}\label{Sec:ConPerspec}
%In this paper, we present infinite-dimensional observers for a large class of boundary controlled port-Hamiltonian systems (BC-PHSs). The input of the proposed observer is such that the error system (between the BC-PHS and the infinite-dimensional observer) is again a BC-PHS. In this way, we are able %to use existing results, from PHS literature, 
%to prove the different convergence of the infinite-dimensional observers.  More specifically,  we are able to show the asymptotic convergence of the observers for a general case of BC-PHS (for any $N\geq 1$), and the exponential convergence for the BC-PHS with first order spatial differential operator ($N = 1$) and with second order ones ($N=2$). The Eurler-Bernoulli beam ($N = 2$) is used as example with two measurements at one end of the beam. The reduction of the observer is part of the future work of the authors as well as the application on observer-based control laws.
A class of infinite-dimensional observer for 1D BC-PHS with differential operators of order $N \geq 1$ and internal damping has been proposed. The convergence of the proposed observer depends on the number and location of available boundary measurements. Provided that enough boundary measurements are available, exponential convergence can be assured for $N=1$ and $N=2$ (Proposition~\ref{Prop:N1Exponential} and \ref{Prop:N2Exponential}) and asymptotic convergence for $N > 1$ (Proposition~\ref{Prop:Asymptotic}). Furthermore, for a class of partitioned BC-PHS i.e. BC-PHS with specific structure,such as the Euler-Bernoulli beam, exponential convergence can be achieved when $N=2$ (Proposition~\ref{Prop:N2ExponentialRelaxed}) and less measurements are available. The Euler-Bernoulli beam model has been used to illustrate the design and numerical performance of the proposed observer. Future work will deal \textcolor{hector}{with stability and performance analysis under the presence of noise and observer-based boundary control.} 

\vspace{-0.5cm}
\bibliography{bibliographie}

\begin{thebibliography}{10}

\bibitem{Maschke1992ConferencePort}
B.~M. Maschke and A.~J. van~der Schaft, ``Port-controlled {H}amiltonian
  systems: modelling origins and systemtheoretic properties,'' {\em IFAC
  Proceedings Volumes}, vol.~25, no.~13, pp.~359--365, 1992.

\bibitem{vanderSchaft2002JournalHamiltonian}
A.~van~der Schaft and B.~M. Maschke, ``{H}amiltonian formulation of
  distributed-parameter systems with boundary energy flow,'' {\em Journal of
  Geometry and Physics}, vol.~42, no.~1-2, pp.~166--194, 2002.

\bibitem{Duindam2009BookModeling}
V.~Duindam, A.~Macchelli, S.~Stramigioli, and H.~Bruyninckx, {\em Modeling and
  control of complex physical systems: the port-{H}amiltonian approach}.
\newblock Springer Science \& Business Media, 2009.

\bibitem{Trenchant2018JournalFinite}
V.~Trenchant, H.~Ramirez, P.~Kotyczka, and Y.~Le~Gorrec, ``{Finite differences
  on staggered grids preserving the port-{H}amiltonian structure with
  application to an acoustic duct},'' {\em Journal of Computational Physics},
  vol.~373, pp.~673--697, November 2018.

\bibitem{Cardoso2020JournalAPartitioned}
F.~L. Cardoso-Ribeiro, D.~Matignon, and L.~Lef{\`e}vre, ``A partitioned finite
  element method for power-preserving discretization of open systems of
  conservation laws,'' {\em IMA Journal of Mathematical Control and
  Information}, vol.~38, no.~2, pp.~493--533, 2021.

\bibitem{LeGorrec2005JournalDirac}
Y.~Le~Gorrec, H.~Zwart, and B.~Maschke, ``Dirac structures and boundary control
  systems associated with skew-symmetric differential operators,'' {\em SIAM
  journal on control and optimization}, vol.~44, no.~5, pp.~1864--1892, 2005.

\bibitem{Jacob2012BookLinear}
B.~Jacob and H.~J. Zwart, {\em Linear port-{H}amiltonian systems on
  infinite-dimensional spaces}, vol.~223.
\newblock Springer Science \& Business Media, 2012.

\bibitem{Augner2014JournalStability}
B.~Augner and B.~Jacob, ``Stability and stabilization of infinite-dimensional
  linear port-{H}amiltonian systems,'' {\em Evolution Equations and Control
  Theory}, vol.~3, no.~2, pp.~207--229, 2014.

\bibitem{Ramirez2014JournalExponential}
H.~Ramirez, Y.~{Le Gorrec}, A.~Macchelli, and H.~Zwart, ``Exponential
  stabilization of boundary controlled port-{H}amiltonian systems with dynamic
  feedback,'' {\em IEEE Transactions on Automatic Control}, vol.~59,
  pp.~2849--2855, Oct 2014.

\bibitem{Macchelli2017JournalOnTheSynthesis}
A.~Macchelli, Y.~{Le Gorrec}, H.~Ramirez, and H.~Zwart, ``On the synthesis of
  boundary control laws for distributed port-{H}amiltonian systems,'' {\em IEEE
  Transactions on Automatic Control}, vol.~62, pp.~1700--1713, April 2017.

\bibitem{Macchelli2020JournalExponential}
A.~Macchelli, Y.~Le~Gorrec, and H.~Ram{\'\i}rez, ``Exponential stabilization of
  port-{H}amiltonian boundary control systems via energy shaping,'' {\em IEEE
  Transactions on Automatic Control}, vol.~65, no.~10, pp.~4440--4447, 2020.

\bibitem{Hidayat2011ConferenceObservers}
Z.~Hidayat, R.~Babuska, B.~De~Schutter, and A.~Nunez, ``Observers for linear
  distributed-parameter systems: A survey,'' in {\em 2011 IEEE international
  symposium on robotic and sensors environments (ROSE)}, pp.~166--171, IEEE,
  2011.

\bibitem{Guo2007JournalTheStabilization}
B.-Z. Guo and C.-Z. Xu, ``The stabilization of a one-dimensional wave equation
  by boundary feedback with noncollocated observation,'' {\em IEEE Transactions
  on Automatic Control}, vol.~52, no.~2, pp.~371--377, 2007.

\bibitem{Krstic2008JournalOutput}
M.~Krstic, B.-Z. Guo, A.~Balogh, and A.~Smyshlyaev, ``Output-feedback
  stabilization of an unstable wave equation,'' {\em Automatica}, vol.~44,
  no.~1, pp.~63--74, 2008.

\bibitem{Guo2009JournalTheStrong}
B.-Z. Guo and W.~Guo, ``The strong stabilization of a one-dimensional wave
  equation by non-collocated dynamic boundary feedback control,'' {\em
  Automatica}, vol.~45, no.~3, pp.~790--797, 2009.

\bibitem{Smyshlyaev2009JournalBoundary}
A.~Smyshlyaev and M.~Krstic, ``Boundary control of an anti-stable wave equation
  with anti-damping on the uncontrolled boundary,'' {\em Systems \& Control
  Letters}, vol.~58, no.~8, pp.~617--623, 2009.

\bibitem{Meurer2011JournalTracking}
T.~Meurer and A.~Kugi, ``Tracking control design for a wave equation with
  dynamic boundary conditions modeling a piezoelectric stack actuator,'' {\em
  International Journal of Robust and Nonlinear Control}, vol.~21, no.~5,
  pp.~542--562, 2011.

\bibitem{Feng2016JournalObserver}
H.~Feng and B.-Z. Guo, ``Observer design and exponential stabilization for wave
  equation in energy space by boundary displacement measurement only,'' {\em
  IEEE Transactions on Automatic Control}, vol.~62, no.~3, pp.~1438--1444,
  2016.

\bibitem{Smyshlyaev2005JournalBackstepping}
A.~Smyshlyaev and M.~Krstic, ``Backstepping observers for a class of parabolic
  {PDEs},'' {\em Systems \& Control Letters}, vol.~54, no.~7, pp.~613--625,
  2005.

\bibitem{Meurer2009JournalTracking}
T.~Meurer and A.~Kugi, ``Tracking control for boundary controlled parabolic
  {PDEs} with varying parameters: Combining backstepping and differential
  flatness,'' {\em Automatica}, vol.~45, no.~5, pp.~1182--1194, 2009.

\bibitem{Toledo2020ConferencePassive}
J.~Toledo, H.~Ramirez, Y.~Wu, and Y.~Le~Gorrec, ``Passive observers for
  distributed port-{H}amiltonian systems,'' in {\em 21st IFAC World Congress,
  2020, July 12-17, 2020, Berlin, Germany}, 2020.

\bibitem{Malzer2020JournalStability}
T.~Malzer, H.~Rams, B.~Kolar, and M.~Sch{\"o}berl, ``Stability analysis of the
  observer error of an in-domain actuated vibrating string,'' {\em IEEE Control
  Systems Letters}, vol.~5, no.~4, pp.~1237--1242, 2020.

\bibitem{Wu2020JournalReduced}
Y.~Wu, B.~Hamroun, Y.~Le~Gorrec, and B.~Maschke, ``Reduced order {LQG} control
  design for infinite dimensional port {H}amiltonian systems,'' {\em IEEE
  Transactions on Automatic Control}, 2020.

\bibitem{Toledo2020JournalObserver}
J.~Toledo, Y.~Wu, H.~Ramirez, and Y.~Le~Gorrec, ``Observer-based boundary
  control of distributed port-{H}amiltonian systems,'' {\em Automatica},
  vol.~120, p.~109130, 2020.

\bibitem{Toledo2022ConferenceObserver}
J.~Toledo, Y.~Wu, H.~Ramirez, and Y.~L. Gorrec, ``Observer design for 1-d
  boundary controlled port-{H}amiltonian systems with different boundary
  measurements,'' {\em IFAC-PapersOnLine}, vol.~55, no.~26, pp.~95--100, 2022.
\newblock 4th IFAC Workshop on Control of Systems Governed by Partial
  Differential Equations CPDE 2022.

\bibitem{Villegas2009JournalExponential}
J.~A. Villegas, H.~Zwart, Y.~Le~Gorrec, and B.~Maschke, ``Exponential stability
  of a class of boundary control systems,'' {\em IEEE Transactions on Automatic
  Control}, vol.~54, no.~1, pp.~142--147, 2009.

\end{thebibliography}
\bibliographystyle{ieeetr}

%\appendix

  % in the appendices.
\end{document}